\documentclass[10pt,twocolumn]{IEEEtran}
\usepackage{amssymb,amsfonts,amsmath}       
\usepackage{algorithm,algorithmic}          
\usepackage{graphicx,subfigure}             
\usepackage{cite}                           
\usepackage{caption}                        

\begin{document}
\captionsetup{font={small}}
\title{\huge UAV-enabled Secure Communication with Finite Blocklength}

\author{Yuntian Wang, Xiaobo Zhou, Zhihong Zhuang, Linlin Sun, Yuwen Qian, Jinhui Lu, and Feng Shu
\thanks{~The authors are with School of Electronic and Optical Engineering, Nanjing University of Science and Technology, Nanjing, 210094, China (e-mail: wytpzaaa@njust.edu.cn; zxb@njust.edu.cn; nustcn@163.com; sunlinlin@mail.njust.edu.cn; admon@njust.edu.cn; lujinhui\_{}njust\_{}1962@163.com; shufeng@njust.edu.cn). (Corresponding authors: Zhihong Zhuang, and Feng Shu)}}

\maketitle

\begin{abstract}
In the finite blocklength scenario, which is suitable for practical applications, a method of maximizing the average effective secrecy rate (AESR) is proposed for a UAV-enabled secure communication by optimizing the UAV's trajectory and transmit power subject to the UAV's mobility constraints and transmit power constraints. To address the formulated non-convex optimization problem, it is first decomposed into two non-convex subproblems. Then the two subproblems are converted respectively into two convex subproblems via the first-order approximation. Finally, an alternating iteration algorithm is developed by solving the two subproblems iteratively using successive convex approximation (SCA) technique. Numerical results show that our proposed scheme achieves a better AESR performance than both the benchmark schemes.
\end{abstract}

\begin{IEEEkeywords}
UAV communications, Physical Layer Security, Trajectory Optimization, Transmit Power Optimization, Finite Blocklength.
\end{IEEEkeywords}

\section{Introduction}
In recent years, unmanned aerial vehicles (UAVs) have attracted extensive attention in the field of wireless communications due to the advantages of low cost, flexible mobility and high probabilities of line of sight (LoS) channels \cite{7470933}. For example, in \cite{8254949}, a UAV was employed as an aerial base station to communicate with the ground users in time division multiple access (TDMA) manner. By optimizing the UAV trajectory and user scheduling, the max-min average rate can be achieved. Besides, the scenario of multi-UAV operating in a cooperation manner was considered in \cite{8247211}. Furthermore, a UAV can also act as a mobile relay when there is no direct link between the transmitter and the receiver \cite{7572068}.

On the other hand, note that wireless communications are vulnerable to wiretaps and attacks due to the nature of broadcast \cite{8290952}. Meanwhile, the potential eavesdropping and attacks on UAV networks are more serious due to the air-to-ground LoS channels. Consequently, the security of UAV networks has been widely investigated. For instance, the authors of \cite{8618602} maximized the secrecy rate by jointly optimizing the UAV's trajectory and transmit power. To further improve the physical layer security (PLS) performance, the authors of \cite{8643815} considered the case of dual UAV, in which one was used to transmit the confidential information (CI) while the other was used to transmit the artificial noise (AN) to prevent eavesdroppers (Eves) from wiretapping. The recent work \cite{8873672} investigated a UAV-enabled mobile edge computing system, where a full-duplex legitimate UAV and non-offloading ground users transmit AN together to cope with the aerial eavesdropping UAVs. Besides, in order to hide the transmission behavior of the transmitter, \cite{8764452} combined a UAV network with covert communications, which can provide high-level security.

Note that the aforementioned works related to PLS all assumed that CI has infinite blocklength. However, this assumption is unrealizable in practice. Especially in UAV networks, where the UAV's flight period is generally discretized into $N$ time slots to facilitate designing the UAV's trajectory, and the duration of each time slot is assumed to be sufficiently small to improve the accuracy of the approximation \cite{8618602,8643815,8873672,8764452}. As such, the assumption of infinite blocklength may affect the validity of the approximation since it is impossible to transmit a packet with infinite blocklength within such a sufficiently small time slot. Additionally, the packet with infinite blocklength can not meet the demands of some applications for 5G wireless systems on ultra-reliable and low-latency communication (URLLC), such as factory automation and autonomous vehicles, which require at least 99.999\% reliability within 1 ms end-to-end latency \cite{8329620,7945856,8329619}. As a result, the packet with finite blocklength will be used to overcome these defects. In this case, the law of large numbers is no longer true, and the thermal noise and the distortions at the receiver can not be averaged out \cite{7529226}. Consequently, the expression of secrecy rate leading to the results of \cite{8618602,8643815,8873672,8764452} can not be used directly \cite{7541867}.

Motivated by the above reasons, in this work, we study a UAV-enabled secure communication with finite blocklength, where a UAV is used to transmit CI with finite blocklength to the legitimate user (Bob) in the existence of Eve. In order to guarantee the communication security of our considered system, we aim to maximize the average effective secrecy rate (AESR) subject to the UAV's mobility and transmit power constraints by jointly optimizing the UAV's trajectory and transmit power. The main contributions of this work are summarized as follows.
\begin{itemize}
  \item For the first time, we consider a secure UAV communication with finite blocklength. The results of information theory on finite blocklength bounds for wiretap channels are utilized to approximate AESR of the considered system, which are quite different from the results of the aforementioned works considering infinite blocklength.
  \item  The formulated optimization problem is non-convex and difficult to solve directly. Therefore, we first decompose it into two non-convex subproblems. Then the two subproblems are transformed into convex ones based on the first-order approximation. Finally, an alternating iteration algorithm based on successive convex approximate (SCA) technique is proposed to solve the formulated problem.
  \item Numerical results show that the blocklength has little effect on the UAV's trajectory, but a great effect on the UAV's transmit power. Besides, the proposed alternating iteration algorithm significantly outperforms both the benchmark schemes from the perspective of AESR.
\end{itemize}

The remainder of this paper is organized as follows: In Section \ref{SMPF}, we present the system model and formulate the optimization problem. Then, an alternating iterative algorithm is proposed in Section \ref{IAP} to solve the formulated optimization problem. Numerical results are provided in Section \ref{NR} to demonstrate the effectiveness of our proposed algorithm. Finally, the conclusions are drawn in Section \ref{conclusion}.

\section{System Model and Problem Formulation}\label{SMPF}
\begin{figure}[!t]
  \centering
  \includegraphics[width=0.35\textwidth]{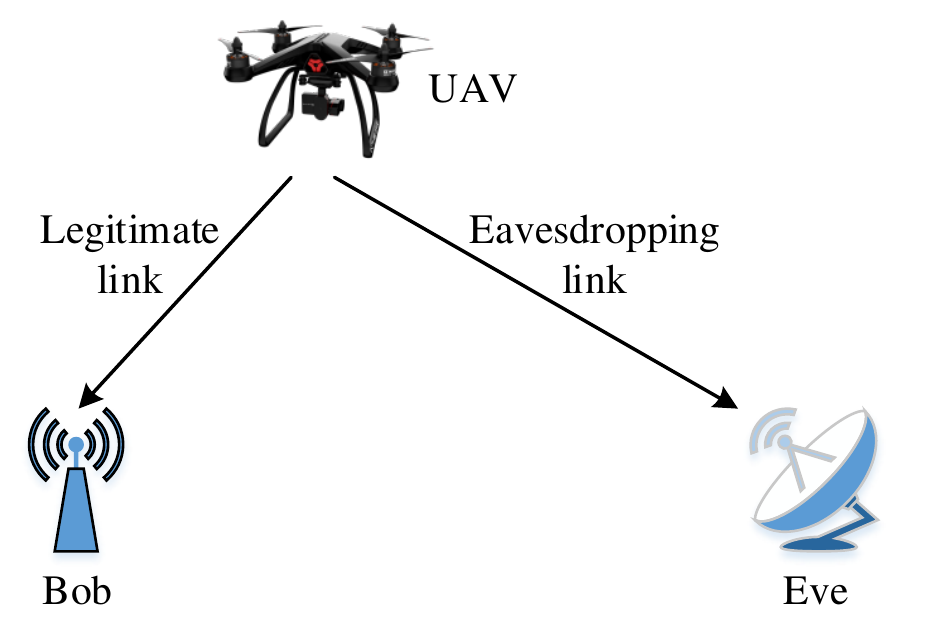}
  \caption{UAV-enabled secure communication system with finite blocklength.}
  \label{model}
\end{figure}
As shown in Fig.~\ref{model}, we consider a UAV-enabled secure communication system, where a UAV serves as an aerial base station to transmit CI to Bob, while a potential Eve tries to wiretap CI. In contrast to the conventional UAV-enabled secure communication systems where infinite blocklength was assumed for CI, we consider the case of CI with finite blocklength. For ease of presentation, we consider a three dimension (3D) Cartesian coordinate system, where Bob and Eve are located at $\mathbf{w}_b=[x_b,y_b,0]^T$ and $\mathbf{w}_e=[x_e,y_e,0]^T$, respectively. The trajectory of UAV within flight period $T$ can be expressed as $\mathbf{q}(t)=[x(t),y(t),H]^T,0\leq t\leq T$, where $H$ denotes the fixed flight altitude. For the sake of simplicity, the flight period $T$ is equally divided into $N$ time slots with duration $\delta_t=\frac{T}{N}$. As a result, the trajectory of UAV can be discretized into a sequence $\mathbf{q}[n]=[x[n],y[n],H]^T,1\leq n\leq N$. Then, the mobility constraints of UAV are given by
\begin{subequations}\label{mobility}
\begin{equation}\label{maximum_speed}
\|\mathbf{q}[n+1]-\mathbf{q}[n]\|\leq V_\mathrm{max}\delta_t,n=1,2,\ldots,N-1,
\end{equation}
\begin{equation}\label{ini_fin_location}
\mathbf{q}[1]=\mathbf{q}_I,~~\mathbf{q}[N]=\mathbf{q}_F,
\end{equation}
\end{subequations}
where $V_\mathrm{max}$ denotes the maximum flight speed of UAV, $\mathbf{q}_I=[x_I,y_I,H]^T$ and $\mathbf{q}_F=[x_F,y_F,H]^T$ denote the predetermined initial and final location of UAV, respectively.

In this work, all the nodes are assumed to be equipped with single antenna. Besides, we assume that the communication links from UAV to Bob and Eve are dominated by LoS channels. As a result, the channel power gains from UAV to Bob and Eve in time slot $n$ can be respectively expressed as
\begin{equation}\label{channel}
h_i[n]=\zeta_0d_b^{-2}[n]=\frac{\zeta_0}{\|\mathbf{q}[n]-\mathbf{w}_b\|^2}, i\in\{b,e\},\forall n,
\end{equation}
where $d_i[n]=\|\mathbf{q}[n]-w_i\|$ denotes the distance from UAV to Bob and Eve in time slot $n$, and $\zeta_0$ denotes the channel power gain at the reference distance $d_0=1$ m. Assume that the transmit power of UAV in time slot $n$ is denoted as $P[n]$. Taking the average and instantaneous power limits into accountant \cite{8618602}, the transmit power constraints are given by
\begin{subequations}\label{power_cons}
\begin{equation}\label{average_power}
\frac{1}{N}\sum_{n=1}^{N}P[n]\leq \bar{P},
\end{equation}
\begin{equation}\label{maximum_power}
0\leq P[n]\leq P_\mathrm{max},\forall n,
\end{equation}
\end{subequations}
where $\bar{P}$ and $P_\mathrm{max}$ denote the maximum average and instantaneous transmit power, respectively. Then, according to \cite{7541867} and \cite{8672179}, a lower bound on the secrecy rate in bits per channel use (BPCU) in time slot $n$ can be approximated as
\begin{equation}\label{secrecy_rate}
\begin{aligned}
R_s[n]=&\Bigg[\log_2(1+\gamma_b[n])-\log_2(1+\gamma_e[n])-\\
&\sqrt{\frac{V_b[n]}{L}}\frac{Q^{-1}(\varepsilon)}{\ln2}-\sqrt{\frac{V_e[n]}{L}}\frac{Q^{-1}(\epsilon)}{\ln2}\Bigg]^+,\forall n,
\end{aligned}
\end{equation}
where the operation $[x]^+\triangleq\max\{x,0\}$, $Q^{-1}(x)$ is the inverse of the Gaussian Q-function $Q(x)\triangleq\int_{x}^{\infty}\frac{1}{\sqrt{2\pi}}e^{-\frac{t^2}{2}}dt$. We note that $L$ in (\ref{secrecy_rate}) denotes the blocklength, $\varepsilon$ and $\epsilon$ denote the decoding error probability at Bob and the information leakage at Eve, respectively. $\gamma_i[n](i\in\{b,e\})$ in (\ref{secrecy_rate}) given by
\begin{equation}\label{SNR_b}
\gamma_i[n]=\frac{P[n]h_i[n]}{\sigma^2}=\frac{\xi_0P[n]}{\|\mathbf{q}[n]-\mathbf{w}_i\|^2}, i\in\{b,e\}, \forall n,
\end{equation}
denote the signal-to-noise ratios (SNRs) at Bob and Eve, where $\sigma^2$ is the additive white Gaussian noise (AWGN) power at Bob and Eve, and $\xi_0=\frac{\zeta_0}{\sigma^2}$. $V_i[n](i\in\{b,e\})$ is the channel dispersion of Bob and Eve in time slot $n$, which is given by \cite{7541867,8672179}
\begin{equation}\label{dispersion_b}
V_i[n]=1-(1+\gamma_i[n])^{-2}, i\in\{b,e\}, \forall n.
\end{equation}

Note that the decoding error probability at Bob cannot be neglected due to finite blocklength. As such, our target is to maximize AESR \cite{8672179} given by $\frac{1}{N}\sum_{n=1}^{N} R_s[n](1-\varepsilon)$ via jointly optimizing the UAV's trajectory and transmit power under the mobility constraints and the transmit power constraints. The resultant optimization problem is formulated as
\begin{subequations}\label{P1}
\begin{equation}\label{objective_P1}
\max_{\mathbf{Q},\mathbf{P}}~\frac{1}{N}\sum_{n=1}^{N} R_s[n](1-\varepsilon)
\end{equation}
\begin{equation}\label{cons_P1}
\text{s.t}.~\text{(\ref{mobility})},\text{(\ref{power_cons})},
\end{equation}
\end{subequations}
where $\mathbf{Q}=\{\mathbf{q}[n],\forall n\}$ is the UAV's trajectory and $\mathbf{P}=\{P[n],\forall n\}$ is the UAV's transmit power over $N$ time slots. In (\ref{objective_P1}), the operation $[x]^+$ is removed because if there exist some time slot $n$ making $R_s[n]<0$, we can always set $P[n]=0$ and then increase $R_s[n]$ to 0 without violating the power constraints (\ref{power_cons}). We note that problem (\ref{P1}) is non-convex and difficult to solve optimally due to the fact that the expression of $R_s$ is very complicated as well as $\mathbf{Q}$ and $\mathbf{P}$ are closely coupled in the objective function. As such, in the next section, we propose an efficient algorithm to find a suboptimal solution to problem (\ref{P1}).

\newcounter{mytempieqncnt}
\begin{figure*}[!t]
\normalsize
\setcounter{mytempieqncnt}{\value{equation}}
\setcounter{equation}{13}
\begin{equation}\label{SCA_objective_P2}
\begin{aligned}
\tilde{R}_s[n]\geq &\log_2\left(1+\frac{\xi_0 P[n]}{\|\mathbf{\hat{q}}[n]-\mathbf{w}_b\|^2}\right)-\frac{\xi_0 P[n](\|\mathbf{q}[n]-\mathbf{w}_b\|^2-\|\mathbf{\hat{q}}[n]-\mathbf{w}_b\|^2)}{(\|\mathbf{\hat{q}}[n]-\mathbf{w}_b\|^2)(\|\mathbf{\hat{q}}[n]-\mathbf{w}_b\|^2+\xi_0 P[n])\ln2}-\\
&\log_2(1+\hat{u}_e[n])-\frac{u_e[n]-\hat{u}_e[n]}{(1+\hat{u}_e[n])\ln2}
-z_b[n]\frac{Q^{-1}(\varepsilon)}{\sqrt{L}\ln2}-z_e[n]\frac{Q^{-1}(\epsilon)}{\sqrt{L}\ln2}\triangleq \tilde{R}_{s,q}^{lb}[n],\forall n.
\end{aligned}
\end{equation}
\setcounter{mytempieqncnt}{\value{equation}}
\setcounter{equation}{16}
\begin{equation}\label{SCA_objective_P4}
\begin{aligned}
\tilde{R}_s[n]\geq&\log_2\left(1\!+\!\frac{\xi_0 P[n]}{\|\mathbf{q}[n]-\mathbf{w}_b\|^2}\right)\!-\!\log_2(1\!+\!\hat{u}_e[n])\!-\!\frac{u_e[n]-\hat{u}_e[n]}{(1+\hat{u}_e[n])\ln2}\!-\!z_b[n]\frac{Q^{-1}(\varepsilon)}{\sqrt{L}\ln2}\!-\!z_e[n]\frac{Q^{-1}(\epsilon)}{\sqrt{L}\ln2}\triangleq \tilde{R}_{s,p}^{lb}[n],\forall n.
\end{aligned}
\end{equation}
\hrulefill
\setcounter{equation}{7}
\end{figure*}

\section{Proposed Alternating Iterative Algorithm} \label{IAP}
To facilitate processing problem (11), we introduce slack variables $\mathbf{U}_i=\{u_i[n],\forall n\}$ and $\mathbf{Z}_i=\{z_i[n],\forall n\}$, $i\in\{b,e\}$, then problem (\ref{P1}) can be reformulated as
\begin{subequations}\label{P1_equ}
\begin{equation}\label{objective_P1_equ}
\max_{\mathbf{Q},\mathbf{P},\mathbf{U}_i,\mathbf{Z}_i}~\frac{1}{N}\sum_{n=1}^{N}\tilde{R}_s[n](1-\varepsilon)
\end{equation}
\begin{equation}\label{cons_P2_equ_1}
\text{s.t}.~\text{(\ref{mobility})},\text{(\ref{power_cons})},
\end{equation}
\begin{equation}\label{cons_P1_equ_2}
u_i[n]\geq \frac{\xi_0P[n]}{\|\mathbf{q}[n]-\mathbf{w}_i\|^2},i\in\{b,e\},\forall n,
\end{equation}
\begin{equation}\label{cons_P1_equ_3}
z_i^2[n]\geq 1-(1+u_i[n])^{-2},i\in\{b,e\},\forall n,
\end{equation}
\begin{equation}\label{cons_P1_equ_4}
z_i[n]\geq 0,i\in\{b,e\},\forall n,
\end{equation}
\end{subequations}
where
\begin{equation}\label{R_s_n}
\begin{aligned}
\tilde{R}_s[n]=\log_2\left(1\!+\!\frac{\xi_0P[n]}{\|\mathbf{q}[n]-\mathbf{w}_b\|^2}\right)\!-\!\log_2(1+u_e[n])-\\
z_b[n]\frac{Q^{-1}(\varepsilon)}{\sqrt{L}\ln2}-z_e[n]\frac{Q^{-1}(\epsilon)}{\sqrt{L}\ln2},\forall n.
\end{aligned}
\end{equation}
We note that problem (\ref{P1_equ}) is equivalent to problem (\ref{P1}) since constraints (\ref{cons_P1_equ_2}) and (\ref{cons_P1_equ_3}) hold with equality at the optimal solution. Otherwise, we can always increase the objective value by decreasing the values of slack variables. However, problem (\ref{P1_equ}) is still non-convex and hard to solve because the objective function (\ref{objective_P1_equ}) is non-concave with respect to (w.r.t.) $\mathbf{Q}$, $\mathbf{P}$ and $\mathbf{U}_e$. Besides, constraints (\ref{cons_P1_equ_2}) and (\ref{cons_P1_equ_3}) are non-convex. As a result, in the following, we first decompose problem (\ref{P1_equ}) into two subproblems, then we transform the two subproblems into convex optimization problems, and finally we develop an alternating iteration algorithm and employ the SCA technique to solve the formulated problem.

\subsection{Trajectory Optimization}
For given feasible transmit power $\mathbf{P}$, problem (\ref{P1_equ}) can be simplified as
\begin{subequations}\label{P2}
\begin{equation}\label{objective_P2}
\max_{\mathbf{Q},\mathbf{U}_i,\mathbf{Z}_i}~\frac{1}{N}\sum_{n=1}^{N}\tilde{R}_s[n](1-\varepsilon)
\end{equation}
\begin{equation}\label{cons_P2_1}
\text{s.t}.~\text{(\ref{mobility})},\text{(\ref{cons_P1_equ_2})}-\text{(\ref{cons_P1_equ_4})}.
\end{equation}
\end{subequations}
However, problem (\ref{P2}) is still non-convex because the objective function (\ref{objective_P2}) is non-concave, and constraints (\ref{cons_P1_equ_2}) and (\ref{cons_P1_equ_3}) are non-convex. In the following, we focus on transforming problem (\ref{P2}) into a convex optimization problem.

First, we introduce slack variable $\mathbf{L}_i=\{l_i[n],\forall n\},i\in\{b,e\}$, and equivalently rewrite constraint (\ref{cons_P1_equ_2}) as
\begin{subequations}\label{cons_P2_2_trans}
\begin{equation}\label{cons_P2_2_1}
u_i[n]\geq \frac{\xi_0P[n]}{l_i[n]},i\in\{b,e\},\forall n,
\end{equation}
\begin{equation}\label{cons_P2_2_2}
\|\mathbf{q}[n]-\mathbf{w}_i\|^2\geq l_i[n],i\in\{b,e\},\forall n.
\end{equation}
\end{subequations}
Note that constraint (\ref{cons_P2_2_1}) is convex now, but constraint (\ref{cons_P2_2_2}) is still non-convex due to the superlevel of a convex function. It is well known that a convex (concave) function is lower (upper) bounded by its first-order Taylor expansion. This motivates us to use the first-order approximation technique to tackle the non-convex constraint (\ref{cons_P2_2_2}). Specifically, the term $\|\mathbf{q}[n]-\mathbf{w}_b\|^2$ can be replaced by its first-order Taylor expansion. Thus, for given feasible point $\mathbf{\hat{q}}[n]$, constraint (\ref{cons_P2_2_2}) can be approximated as
\begin{equation}\label{SCA_cons_P2_2}
\|\mathbf{\hat{q}}[n]\!-\!\mathbf{w}_i\|^2\!+\!2(\mathbf{\hat{q}}[n]\!-\!\mathbf{w}_i)^T(\mathbf{q}[n]\!-\!\mathbf{\hat{q}}[n])\!\geq\! l_i[n],i\in\{b,e\},\forall n,
\end{equation}
which is now a convex constraint.

For constraint (\ref{cons_P1_equ_3}), we observe that it is in the form of a superlevel of a convex function, which can be approximated by its first-order convex approximation. Consequently, for given feasible points $\hat{z}_i[n]$ and $\hat{u}_i[n]$, constraint (\ref{cons_P1_equ_3}) can be approximated as
\begin{equation}\label{SCA_cons_P1_equ_3}
\begin{aligned}
\hat{z}_i^2[n]+2\hat{z}_i[n](z_i[n]-\hat{z}_i[n])\geq 1-(1+\hat{u}_i[n])^{-2}\\
+2(1+\hat{u}_i[n])^{-3}(u_i[n]-\hat{u}_i[n]),i\in\{b,e\},\forall n.
\end{aligned}
\end{equation}

For the objective function (\ref{objective_P2}), we observe that $\tilde{R}_s[n]$ in the objective function is jointly convex w.r.t. $\|\mathbf{q}[n]-\mathbf{w}_b\|^2$ and $u_e[n]$. Consequently, the first-order approximation technique can be used to construct a lower bound of $\tilde{R}_s[n]$, which can be expressed as (\ref{SCA_objective_P2}) on the top of this page.

Following the above transformation, problem (\ref{P2}) can be reformulated as
\setcounter{equation}{14}
\begin{subequations}\label{P3}
\begin{equation}\label{objective_P3}
\max_{\mathbf{Q},\mathbf{U}_i,\mathbf{Z}_i,\mathbf{L}_i}~\frac{1}{N}\sum_{n=1}^{N}\tilde{R}_{s,q}^{lb}[n](1-\varepsilon)
\end{equation}
\begin{equation}\label{cons_P3_1}
\begin{aligned}
\text{s.t}.~\text{(\ref{mobility})},\text{(\ref{cons_P1_equ_4})},\text{(\ref{cons_P2_2_1})},\text{(\ref{SCA_cons_P2_2})},\text{(\ref{SCA_cons_P1_equ_3})}.
\end{aligned}
\end{equation}
\end{subequations}
Problem (\ref{P3}) is a convex optimization problem, which can be efficiently solved by optimization tools such as CVX \cite{cvx}.

\subsection{Transmit Power Optimization}
For given feasible UAV's trajectory $\mathbf{Q}$, problem (\ref{P1_equ}) can be simplified as
\begin{subequations}\label{P4}
\begin{equation}\label{objective_P4}
\max_{\mathbf{P},\mathbf{U}_i,\mathbf{Z}_i}~\frac{1}{N}\sum_{n=1}^{N}\tilde{R}_s[n](1-\varepsilon)
\end{equation}
\begin{equation}\label{cons_P4_1}
\text{s.t}.~\text{(\ref{power_cons})},\text{(\ref{cons_P1_equ_2})}-\text{(\ref{cons_P1_equ_4})}.
\end{equation}
\end{subequations}
Obviously, problem (\ref{P4}) is non-convex due to the convexity of the second term of $\tilde{R}_s[n]$ and the non-convexity of constraints (\ref{cons_P1_equ_3}). Similar to the process in the previous subsection, constraints (\ref{cons_P1_equ_3}) can be approximated as constraint (\ref{SCA_cons_P1_equ_3}). For the objective function (\ref{objective_P4}), we can employ the first-order approximation to construct a lower bound of $\tilde{R}_s[n]$, which is detailed as (\ref{SCA_objective_P4}) on the top of last page.

Then, problem (\ref{P4}) can be approximated as
\setcounter{equation}{17}
\begin{subequations}\label{P5}
\begin{equation}\label{objective_P5}
\max_{\substack{\mathbf{P},\mathbf{U}_i, \mathbf{Z}_i}}~\frac{1}{N}\sum_{n=1}^{N}\tilde{R}_{s,p}^{lb}[n](1-\varepsilon)
\end{equation}
\begin{equation}\label{cons_P5_1}
\text{s.t}.~\text{(\ref{power_cons})},\text{(\ref{cons_P1_equ_2})},\text{(\ref{cons_P1_equ_4})},\text{(\ref{SCA_cons_P1_equ_3})}.
\end{equation}
\end{subequations}
We note that problem (\ref{P5}) is a convex optimization problem, which can be efficiently solved by CVX \cite{cvx}.

\begin{algorithm}[!t]
\caption{Alternating iterative algorithm for problem (\ref{P1})}
\label{AL1}
\begin{algorithmic}[1]
\STATE Initialize \{$\mathbf{Q}^0$, $\mathbf{P}^0$, $\mathbf{U}_i^0$, $\mathbf{Z}_i^0$\}; Let $r=0$.
\REPEAT
\STATE Solve problem (\ref{P3}) for given \{$\mathbf{Q}^r$, $\mathbf{P}^r$, $\mathbf{U}_i^r$, $\mathbf{Z}_i^r$\} and obtain the optimal solution \{$\mathbf{Q}^{r+1}$, $\mathbf{U}_i^{r+1}$, $\mathbf{Z}_i^{r+1}$\}.
\STATE Let $\{\mathbf{U}_i^{r}, \mathbf{Z}_i^r\}\!=\!\{\mathbf{U}_i^{r+1}, \mathbf{Z}_i^{r+1}\}$.
\STATE Solve problem (\ref{P5}) for given \{$\mathbf{Q}^{r+1}$, $\mathbf{U}_i^r$, $\mathbf{Z}_i^r$\} and obtain the optimal solution \{$\mathbf{P}^{r+1}$, $\mathbf{U}_i^{r+1}$, $\mathbf{Z}_i^{r+1}$\}.
\STATE Let $r=r+1$.
\UNTIL{the fractional increase of the objective function is below a threshold $\tau$.}
\end{algorithmic}
\end{algorithm}

\subsection{Overall Algorithm}
In the previous two subsections, we have transformed the trajectory optimization subproblem and the transmit power optimization subproblem into convex optimization problems. In this subsection, we develop an alternating iteration algorithm based on the SCA technique to solve the two subproblems alternatively. Following the principle of the SCA technique, at each iteration, the current optimal solution to each subproblem gradually approximates the solution to the original optimization problem (\ref{P1}). Furthermore, the optimal solution to each subproblem is also feasible to problem (\ref{P1}), since the feasible set of each subproblem is stricter than that of problem (\ref{P1}). As such, the proposed algorithm can obtain a suboptimal solution to the original optimization problem (\ref{P1}). The detailed algorithm is shown in Algorithm \ref{AL1}. As per \cite{8247211}, Algorithm \ref{AL1} is guaranteed to converge since the objective value of problem (\ref{P1}) is non-decreasing at each iteration and the objective value of problem (\ref{P1}) is bounded.

\section{Numerical Results} \label{NR}
In this section, we present numerical results to demonstrate the performance of our proposed algorithm. The simulation parameters are set as: $P_\mathrm{max}=20$ dBm, $\bar{P}=\frac{1}{2}P_\mathrm{max}$, $H=100$ m, $\delta_t=1$ s, $V_\mathrm{max}=10$ m/s, $L=400$, $\xi_0=60$ dB, $\varepsilon=10^{-5}$, $\epsilon=10^{-2}$, $\tau=10^{-6}$, $\mathbf{w}_b=[0,0,0]^T$ m, $\mathbf{w}_e=[400,0,0]^T$ m, $\mathbf{q}_I=[200,100,H]^T$ m and $\mathbf{q}_F=[200,-100,H]^T$ m. For ease of presentation, the proposed algorithm, i.e., joint optimization of the UAV's trajectory and transmit power, is denoted as the JTPO scheme. For comparison, we consider the following two benchmark schemes:
\begin{itemize}
  \item FTP-Inf: Fixed trajectory and transmit power under the case of infinite blocklength, where the trajectory and transmit power are obtained by the method in \cite{8618602}.
  \item POFT: Transmit power optimization with fixed trajectory, where the transmit power is obtained by solving problem (\ref{P5}), and the line segment trajectory in \cite{8764452} is adopted.
\end{itemize}

\begin{figure}[!t]
  \centering
  \includegraphics[width=0.4\textwidth]{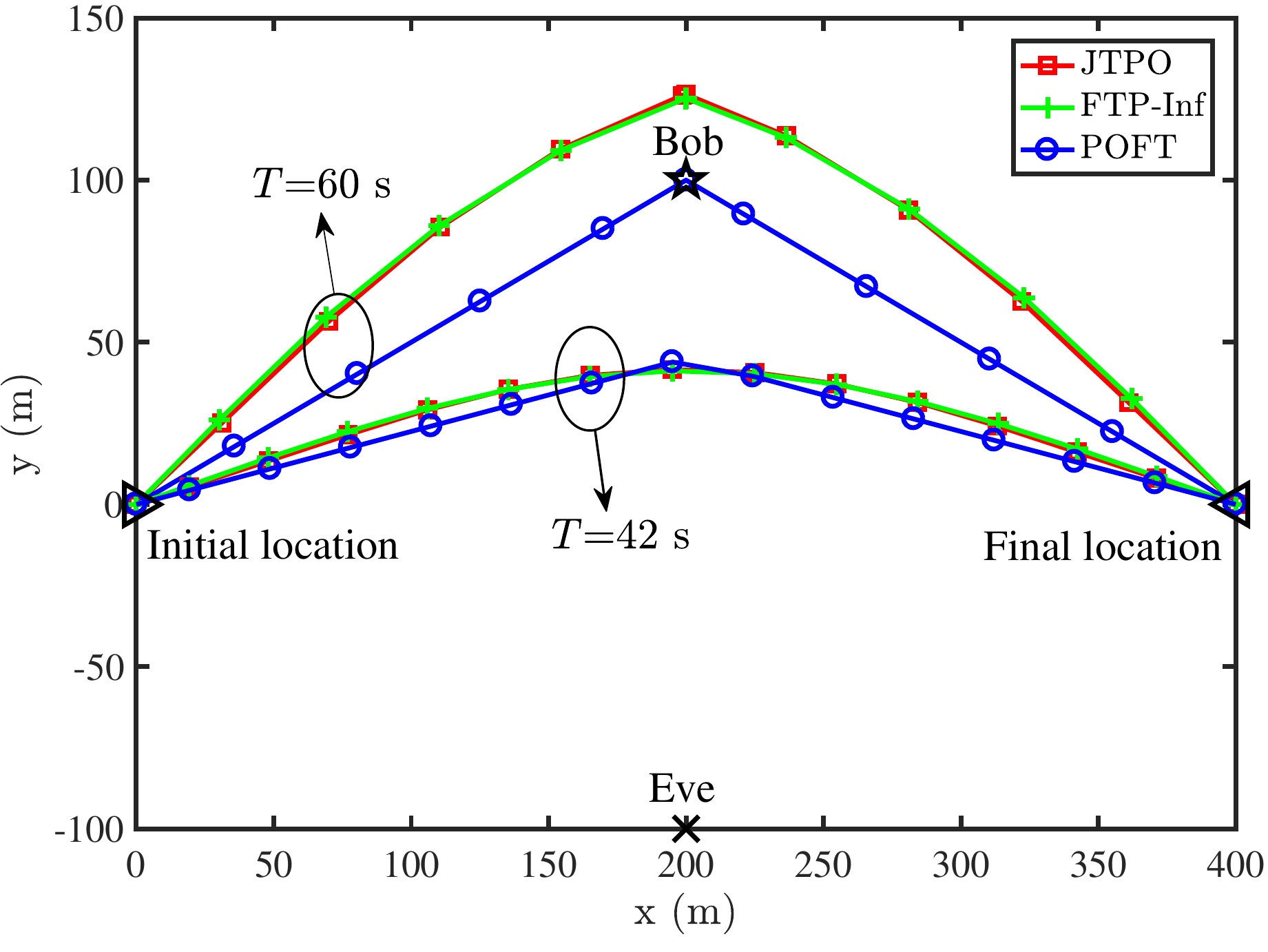}
  \caption{UAV’s trajectories versus different values of the flight period $T$.}
  \label{trajectory}
\end{figure}

\begin{figure}[!t]
  \centering
  \includegraphics[width=0.5\textwidth]{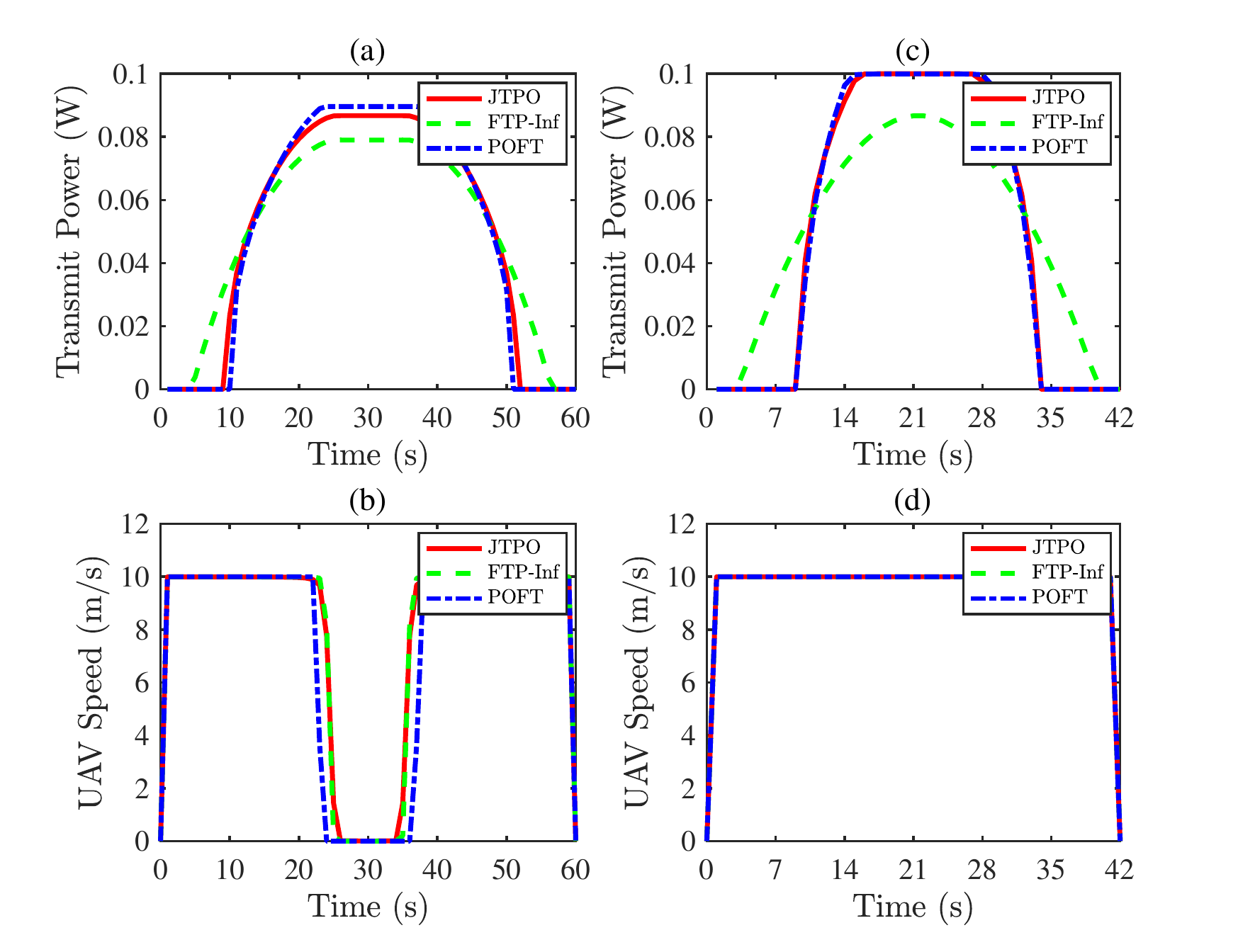}
  \caption{UAV’s transmit power and speed for different values of the flight period $T$, where $T=60$ s for (a) and (b), while $T=42$ s for (c) and (d).}
  \label{compare}
\end{figure}

Fig.~\ref{trajectory} plots the UAV trajectories obtained by the JTPO and the two benchmark schemes versus different values of the flight period $T$. In this figure, we observe that the UAV trajectories obtained by the JTPO scheme are similar to those obtained by the FTP-Inf scheme. Specifically, when $T=60$ s, the UAV first flies towards a certain location around Bob at the maximum speed, then hovers there as long as possible, and finally heads for the final location at the maximum speed. This observation is demonstrated by the UAV’s speed shown in Fig.~\ref{compare}(b). We also note that the hovering location is rightly above Bob while the eavesdropper locates rightly below Bob. As such, the hovering location is a tradoff between the communication performance and the security performance. Compared with the line segment trajectory obtained by the POFT scheme, the trajectories obtained by the JTPO and FTP-Inf schemes always fly farther away from Eve to deteriorate Eve’s eavesdropping. When $T=42$ s, the flight period $T$ is not large enough for the UAV to reach the location of Bob. Consequently, as confirmed by the UAV's speed given in Fig.~\ref{compare}(d), the UAV flies at the maximum speed from the initial location to the final location via a curved path without hovering to get as close to Bob as possible.

Fig.~\ref{compare}(a) and Fig.~\ref{compare}(c) present the UAV's transmit power for different values of the flight period $T$. We observe that Fig. \ref{compare}(a) and Fig.~\ref{compare}(c) are symmetrical. Thus, we only analyze the results of the first half of time slots. It is first observed that the UAV's transmit power obtained by the JTPO and FTP-Inf schemes is quite different, although the UAV's trajectories obtained by these two schemes are similar. Compared with the FTP-Inf scheme, the JTPO and the POFT schemes start transmitting CM later. This is due to the fact that compared with the infinite blocklength, the penalty terms (i.e., the last two terms) on the achievable secrecy rate with finite blocklength make a positive secrecy rate more difficult to be guaranteed. Besides, we can observe that all the three schemes enhance the UAV's transmit power when the UAV flies closer to Bob and lower the UAV's transmit power when the UAV moves farther away from Bob.

\begin{figure}[!t]
  \centering
  \includegraphics[width=0.4\textwidth]{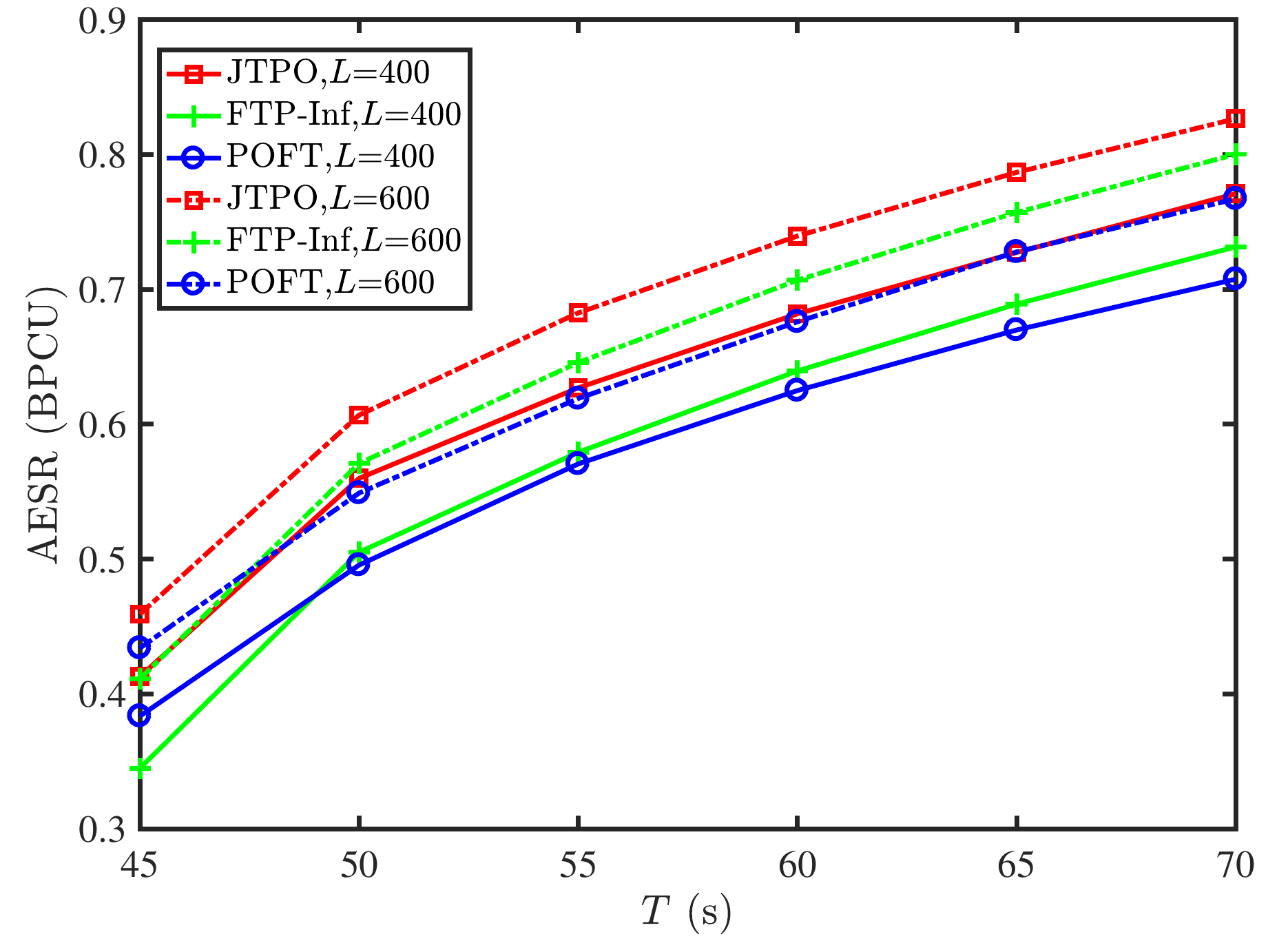}
  \caption{AESR versus the flight period $T$ for different values of the blocklength $L$.}
  \label{Rs_T}
\end{figure}

In Fig.~\ref{Rs_T}, we plot the curves of AESR versus the flight period $T$ for different values of the blocklength $L$. As expected, the proposed JTPO scheme outperforms the FTP-Inf scheme, which demonstrates the necessity of considering the blocklength in optimization. Besides, the proposed JTPO scheme achieves better performance than the POFT scheme thanks to the trajectory optimization. We can also observe that AESRs of all the three schemes rise as $T$ increases since a larger $T$ allows the UAV to hover at the hovering location for a longer time. Finally, AESRs obtained by all the three schemes increase as $L$ increases, which is consistent with the result of (\ref{secrecy_rate}).

\section{Conclusions} \label{conclusion}
In this paper, we have investigated a UAV-enabled secure communication with finite blocklength. Our objective is to maximize AESR by jointly designing the UAV's trajectory and transmit power subject to the UAV’s mobility constraints and transmit power constraints. However, the resultant optimization problem is non-convex and difficult to solve directly. As such, we first decomposed the formulated optimization problem into the trajectory optimization and the transmit power optimization subproblems. Then we reformulated the two subproblems into convex optimization problems based on the first-order approximation technique. Finally, we developed an alternating iteration algorithm based on the SCA technique to solve the two subproblems iteratively. Numerical results showed that the proposed joint optimization scheme can achieve significantly better security performance than both the benchmark schemes.

\bibliographystyle{IEEEtran}
\bibliography{cite.bib}

\begin{thebibliography}{10}
\providecommand{\url}[1]{#1}
\csname url@samestyle\endcsname
\providecommand{\newblock}{\relax}
\providecommand{\bibinfo}[2]{#2}
\providecommand{\BIBentrySTDinterwordspacing}{\spaceskip=0pt\relax}
\providecommand{\BIBentryALTinterwordstretchfactor}{4}
\providecommand{\BIBentryALTinterwordspacing}{\spaceskip=\fontdimen2\font plus
\BIBentryALTinterwordstretchfactor\fontdimen3\font minus
  \fontdimen4\font\relax}
\providecommand{\BIBforeignlanguage}[2]{{%
\expandafter\ifx\csname l@#1\endcsname\relax
\typeout{** WARNING: IEEEtran.bst: No hyphenation pattern has been}%
\typeout{** loaded for the language `#1'. Using the pattern for}%
\typeout{** the default language instead.}%
\else
\language=\csname l@#1\endcsname
\fi
#2}}
\providecommand{\BIBdecl}{\relax}
\BIBdecl

\bibitem{7470933}
Y.~{Zeng}, R.~{Zhang}, and T.~J. {Lim}, ``Wireless communications with unmanned
  aerial vehicles: opportunities and challenges,'' \emph{IEEE Commun. Mag.},
  vol.~54, no.~5, pp. 36--42, May 2016.

\bibitem{8254949}
Q.~{Wu}, Y.~{Zeng}, and R.~{Zhang}, ``Joint trajectory and communication design
  for {UAV}-enabled multiple access,'' in \emph{IEEE Global Commun. Conf.},
  Dec. 2017, pp. 1--6.

\bibitem{8247211}
------, ``Joint trajectory and communication design for multi-{UAV} enabled
  wireless networks,'' \emph{IEEE Trans. Wireless Commun.}, vol.~17, no.~3, pp.
  2109--2121, Mar. 2018.

\bibitem{7572068}
Y.~{Zeng}, R.~{Zhang}, and T.~J. {Lim}, ``Throughput maximization for
  {UAV}-enabled mobile relaying systems,'' \emph{IEEE Trans. Commun.}, vol.~64,
  no.~12, pp. 4983--4996, Dec. 2016.

\bibitem{8290952}
F.~{Shu}, Y.~{Qin}, T.~{Liu}, L.~{Gui}, Y.~{Zhang}, J.~{Li}, and Z.~{Han},
  ``Low-complexity and high-resolution {DOA} estimation for hybrid analog and
  digital massive {MIMO} receive array,'' \emph{IEEE Trans. Commun.}, vol.~66,
  no.~6, pp. 2487--2501, Jun. 2018.

\bibitem{8618602}
G.~{Zhang}, Q.~{Wu}, M.~{Cui}, and R.~{Zhang}, ``Securing {UAV} communications
  via joint trajectory and power control,'' \emph{IEEE Trans. Wireless
  Commun.}, vol.~18, no.~2, pp. 1376--1389, Feb. 2019.

\bibitem{8643815}
X.~{Zhou}, Q.~{Wu}, S.~{Yan}, F.~{Shu}, and J.~{Li}, ``{UAV}-enabled secure
  communications: Joint trajectory and transmit power optimization,''
  \emph{IEEE Trans. Veh. Technol.}, vol.~68, no.~4, pp. 4069--4073, Apr. 2019.

\bibitem{8873672}
Y.~{Zhou}, C.~{Pan}, P.~L. {Yeoh}, K.~{Wang}, M.~{Elkashlan}, B.~{Vucetic}, and
  Y.~{Li}, ``Secure communications for {UAV}-{E}nabled mobile edge computing
  systems,'' \emph{IEEE Trans. Commun.}, vol.~68, no.~1, pp. 376--388, Jan.
  2020.

\bibitem{8764452}
X.~{Zhou}, S.~{Yan}, J.~{Hu}, J.~{Sun}, J.~{Li}, and F.~{Shu}, ``Joint
  optimization of a {UAV}'s trajectory and transmit power for covert
  communications,'' \emph{IEEE Trans. Signal Process.}, vol.~67, no.~16, pp.
  4276--4290, Aug. 2019.

\bibitem{8329620}
J.~{Sachs}, G.~{Wikstrom}, T.~{Dudda}, R.~{Baldemair}, and K.~{Kittichokechai},
  ``5{G} radio network design for ultra-reliable low-latency communication,''
  \emph{IEEE Network}, vol.~32, no.~2, pp. 24--31, Mar. 2018.

\bibitem{7945856}
C.~{She}, C.~{Yang}, and T.~Q.~S. {Quek}, ``Radio resource management for
  ultra-reliable and low-latency communications,'' \emph{IEEE Commun. Mag.},
  vol.~55, no.~6, pp. 72--78, Jun. 2017.

\bibitem{8329619}
P.~{Popovski}, J.~J. {Nielsen}, C.~{Stefanovic}, E.~d.~{Carvalho}, E.~{Strom},
  K.~F. {Trillingsgaard}, A.~{Bana}, D.~M. {Kim}, R.~{Kotaba}, J.~{Park}, and
  R.~B. {Sorensen}, ``Wireless access for ultra-reliable low-latency
  communication: Principles and building blocks,'' \emph{IEEE Network},
  vol.~32, no.~2, pp. 16--23, Mar. 2018.

\bibitem{7529226}
G.~{Durisi}, T.~{Koch}, and P.~{Popovski}, ``Toward massive, ultrareliable, and
  low-latency wireless communication with short packets,'' in \emph{Proc. of
  IEEE}, vol. 104, no.~9, Sep. 2016, pp. 1711--1726.

\bibitem{7541867}
W.~{Yang}, R.~F. {Schaefer}, and H.~V. {Poor}, ``Finite-blocklength bounds for
  wiretap channels,'' in \emph{Proc. IEEE Int. Symp. Inf. Theory}, Jul. 2016,
  pp. 3087--3091.

\bibitem{8672179}
H.~{Wang}, Q.~{Yang}, Z.~{Ding}, and H.~V. {Poor}, ``Secure short-packet
  communications for mission-critical {IoT} applications,'' \emph{IEEE Trans.
  Wireless Commun.}, vol.~18, no.~5, pp. 2565--2578, May 2019.

\bibitem{cvx}
M.~Grant and S.~Boyd, ``{CVX}: Matlab software for disciplined convex
  programming, version 2.1,'' \url{http://cvxr.com/cvx}, Mar. 2014.

\end{thebibliography}
\end{document}